\documentclass[sigconf]{acmart}
\usepackage{booktabs} % For formal tables
\usepackage{tabularx}
\usepackage{todonotes}
\usepackage{balance}
\def\BibTeX{{\rm B\kern-.05em{\sc i\kern-.025em b}\kern-.08emT\kern-.1667em\lower.7ex\hbox{E}\kern-.125emX}}
    
\copyrightyear{2020}
\acmYear{2020}
\setcopyright{acmlicensed}
\acmConference[CHIIR '20]{2020 Conference on Human Information Interaction
and Retrieval}{March 14--18, 2020}{Vancouver, BC, Canada}
\acmBooktitle{2020 Conference on Human Information Interaction and
Retrieval (CHIIR '20), March 14--18, 2020, Vancouver, BC, Canada}
\acmPrice{15.00}
\acmDOI{10.1145/3343413.3378013}
\acmISBN{978-1-4503-6892-6/20/03}

\begin{document}

\fancyhead{}
  % do not delete this code.

% The "title" command has an optional parameter, allowing the author to define a "short title" to be used in page headers.
\title{Would You Like to Hear the News? Investigating Voice-Based Suggestions for Conversational News Recommendation}

% The "author" command and its associated commands are used to define the authors and their affiliations.
% Of note is the shared affiliation of the first two authors, and the "authornote" and "authornotemark" commands
% used to denote shared contribution to the research.
\author{Harshita Sahijwani}
\affiliation{%
  \institution{Emory University}
  \city{Atlanta}
  \state{Georgia}
}
\email{hsahijw@emory.edu}

\author{Jason Ingyu Choi}
\affiliation{%
  \institution{Emory University}
  \city{Atlanta}
  \state{Georgia}
}
\email{in.gyu.choi@emory.edu}

\author{Eugene Agichtein}
\affiliation{%
  \institution{Emory University}
  \city{Atlanta}
  \state{Georgia}
}
\email{yagicht@emory.edu}

%
% The abstract is a short summary of the work to be presented in the article.
\begin{abstract}
One of the key benefits of voice-based personal assistants is the potential to proactively recommend relevant and interesting information. One of the most valuable sources of such information is the News. However, in order for the user to hear the news that is useful and relevant to them, it must be recommended in an interesting and informative way. However, to the best of our knowledge, {\em how} to present a news item for a voice-based recommendation remains an open question. In this paper, we empirically compare different ways of recommending news, or specific news items, in a voice-based conversational setting. 
Specifically, we study the user engagement and satisfaction with five different variants of presenting news recommendations: (1) a generic news briefing; (2) news about a specific entity relevant to the current conversation; (3) news about an entity from a past conversation; (4) news on a trending news topic; and (5) the default - a suggestion to talk about news in general. Our results show that entity-based news recommendations exhibit 29\% higher acceptance compared to briefing recommendations, and almost 100\% higher acceptance compared to recommending generic or trending news. Our investigation into the presentation of news recommendations and the resulting insights could make voice assistants more informative and engaging.
\end{abstract}

\maketitle

\section{Introduction}
\label{sec:introduction}

% \todo[inline]{Follow the argument from the abstract: personal assistants useful if can proactively recommend info, news is one important source, we study how to recommend news. Then describe the example}
Voice-based personal assistants can serve as a medium for getting useful and interesting information in a timely and convenient manner. 
One of their key benefits is their ability to proactively update the user about new political developments, sports events, weather, and a variety of other topics that might be relevant to them.
News is one of the most important sources of useful information. However, in order for the user to hear the news relevant to them, it must be recommended in an interesting way.

An instance of such a recommendation can be seen in the example conversation between a user and an open-domain voice-based conversational agent  in Table \ref{tab:example_conversation}. 
The user talks to the agent about a recent movie, and when the user loses interest in discussing movies, the system recommends a trending technology story to engage the user in hearing the news.

% There has been some work on ranking news events by relevance for initiating news-related small talk in social dialogue systems ~\citep{bechberger2016personalized}.  However,  in an open-domain conversational bot, it is hard to infer enough information about the user's news preferences with high confidence to recommend news based on events. 
%While there has  been some work on ranking events providing personalized news in dialogue systems, for example \cite{bechberger2016personalized} where a learning-to-rank approach for ranking news events using features capturing the user's general interests has been described.  
%  and more generally, on selecting the next topic of conversation~\citep{yan2018smarter}\citep{yan2017joint}. 
\begin{table}[h!]
    \centering
    \begin{tabular}{l l}
        Bot : & Hey, did you see the recent movie Joker? \\
        User: & Yes, I loved it. It's pretty awesome. \\
        Bot : & That's nice. I'm guessing you like superhero movies? \\
        User: & Not really, but Joker is just on another level. Loved the \\
        & character and the aesthetic. \\
        Bot : & I see. Would you like to talk about other movies like \\
        & the Joker? \\
        User: & I don't want to talk about movies. \\
        Bot : & There is a trending technology story about \\
        &Apple's new AirPods Pro. Want to hear it? \\
        User: & Yes.\\
        Bot : &  Wired on Oct 28th reported: Apple revealed a new \\
        & version of AirPods on its website today, adding the \\
        & totally wire-free headphones to its list of products. \\
        & I have more news on this topic. Would you like to hear?
    \end{tabular}
    \caption{Example human-machine conversation. The conversational agent recommends news as the next topic by proposing a trending technology news. (Note: This is not a conversation from an Alexa user.)}
    \label{tab:example_conversation}
\end{table}
\vspace{-8mm}
Our work is complementary to the task of {\em which} item to recommend. Instead, we study an important problem of {\em how} to present a news item in a voice-based conversational setting. 
To answer this question, we experimented with different ways of suggesting news as the next topic to initiate a conversation about news in a large scale user study with live Alexa users.  
In particular, we study the user engagement and satisfaction with five different methods of presenting news recommendations: (1) offering a generic news briefing; (2) proposing to talk about an entity  from the current conversation; (3) proposing to talk about an entity from the user's past conversations; (4) offering news on a trending news topic; and (5) a generic suggestion to talk about news without a specific entity or topic.

% In a large scale user study conducted with thousands of Alexa users, we found that the {\em formulation} of the topic suggestion significantly impacted that amount of interest in the topic elicited in the user.
Our results show that entity-based news recommendations have 29\% higher acceptance compared to briefing recommendations, and almost 100\% higher acceptance compared to recommending trending news. Furthermore, we found that recommending news on an entity that the user has shown interest in, leads to significantly higher likelihood of the user hearing the news. In summary, our contributions include:

%\todo[inline]{Fill in final main results here.}

\begin{table*}[htb]
    \centering
    \begin{tabular}{ p{2.5cm} |p{7.5cm} | p{7.2cm} }
        \bottomrule
         \textbf{Type} & \textbf{Description} & \textbf{Examples}\\
        \bottomrule
       Generic  &  Proposing to talk about news without a specific entity/headline. & "There is so much going on in the world. Would you like to talk about the news? \\
        \bottomrule
       Trending News & Proposing to tell news on a trending topic from Bing News/Google Trends. & "How about a recent sports story titled "Astros solidify comeback". Want to hear it?"\\
        & & "There is a trending technology story about Apple AirPods Pro. Want to hear it?"\\
        \bottomrule
            % \todo[inline]{Rename ``overview'' to ``News Briefing'' everywhere}
       News Briefing & Proposing to provide a 45-second summary of the news from Reuters. & "I prepared a 45-second overview of the news for you. Want to hear it?"\\
        \bottomrule
        Entity based (Current Conversation)& Proposing to tell news about the user's favorite sports team (if they mentioned it earlier), or any entity that the user asked a question about. (How old is the Dalai Lama?) & "Do you want to hear the latest sports updates on your favorite team?"\\
        \bottomrule
        Entity based (Past Conversation) & Proposing to tell news about an entity mentioned by the user in past conversations. & "Last time we spoke, you mentioned Imagine Dragons. Would you like to hear the latest news on Imagine Dragons?" \\
        \bottomrule
    \end{tabular}
    \caption{Five different news recommendation types defined after analyzing common news-related queries issued to our system.}

    \label{tab:recommendation types}
\end{table*}

% Conversational news recommendation is different from web-based news recommendation, because it is not possible to present too many options to the user on a voice interface. It is also important to deliver the content in a format that is understandable and engaging.

\begin{itemize}
\item Collating the different options for presenting news recommendations in an open-domain conversational agent.
\item Defining the evaluation metrics for quantifying the effectiveness of different presentation conditions for conversational news recommendation.
\item Reporting on a large-scale empirical evaluation of conversational news recommendation with real users. 
\end{itemize}

To the best of our knowledge, this paper is the first large-scale, empirical evaluation of different variations of presenting news recommendation in a conversational setting. As our results show, the presentation of the news recommendation has a significant effect on recommendation acceptance, and some effect on subsequent user satisfaction with the conversational system, suggesting promising directions for future work.

% The rest of the paper is organized as follows:
The next section briefly talks about the related work to place our contributions in context.
% Section \ref{sec: system description} describes our news search and ranking methods. 
% Section \ref{sec:types of recommendation} describes the different types of news recommendations offered to the users. 
% Section \ref{sec:experiments} describes the experimental settings used to evaluate the news presentation methods
% Section \ref{sec: results} presents an analysis of the acceptance rates of all the types of recommendation, and explains the motivation behind entity-based news recommendation.

% Section \ref{sec:discussion} contains discussion and future work.
\vspace{-3mm}
\section{Related work}
In this section, we first present an overview of previous literature on general news recommendation. Then, we present approaches related to conversational news recommendation, followed by work on understanding user behaviors to evaluate conversational systems.
\label{sec:background}
\vspace{-3mm}
\subsection{News Recommendation}
Recommendation systems have a long tradition of literature since these systems are widely used in popular applications like e-commerce, video and streaming platforms, and web search. For web-based news recommendation systems like Google News and Bing web search portal, collaborative filtering \citep{das2007google, lu2015content} and hybrid methods \citep{liu2010personalized} are proposed to profile users based on their clicks and offer personalized recommendations. 

% \subsection{Open Domain Conversational Agents}

% Conversational systems like Amazon's Alexa and Apple's Siri have evolved rapidly and become ubiquitous. Dialogue management systems have become more sophisticated, transitioning from rule-based \citep{form-DM, ravenclaw} to end-to-end systems \citep{DM_end, luo2018learning, dhingra2016towards}. To enable emotionally engaging responses, there have been attempts to analyze the effect of dynamic prosody modifications in conversational quality \citep{gunrock_ssml}. Significant advances have been made in the area of automatic speech recognition (ASR) \citep{chiu2018state, speech_RNN} as well as on designing a flexible and scalable architecture for conversational systems \citep{gunrock, sounding_board, alquist}.

% However, despite the work on these essential components of conversational systems, effective recommendation methods for the conversational setting and strategies for their evaluation are less explored. 
% In general, compared to recommender systems in web-based platforms, conversational recommendation in open-domain setting is much more challenging because of cold start problems and subjectivity in conversations. 
% \todo[inline]{say something about why recommendations in conversational setting are different from those on the web} harshita
\vspace{-3mm}
\subsection{News Dialogue Systems}
Yoshino et al. \citep{yoshino2015news} proposed a conversational news navigation system, which offers news headlines one by one, and reads out the full news article or answers questions about it depending on the user query. However, this kind of presentation is not feasible in an open-domain conversational agent.

From our examination of logs of user conversations with our system, we found that all news-related queries belonged to one of the following main categories:

\begin{enumerate}
\item News from a particular category: "Tell me the latest sports news."
\item News about a specific entity: "Tell me some news about the New England Patriots."
\item A news briefing: "Give me my news briefing."
\item Generic: "Let's talk about the news."
\end{enumerate}

In addition, we observed that returning users commonly mention entities from their past conversations.
Based on these observations, we defined five different ways of presenting news recommendations in an open-domain conversational agent, summarized in Table \ref{tab:recommendation types}.

\subsection{Measuring Satisfaction and Engagement in Conversational Agents}
% User behaviors in conversational systems are also very different from well-known metrics in traditional information retrieval systems, such as click-through, dwell-times and touch features \citep{pred, fox2005evaluating}. 
Traditional IR systems have been evaluated by studying user behaviors using well-known metrics such as click-through, dwell-times and touch features \citep{pred, fox2005evaluating}.
% There have been successful attempts to evaluate conversational systems based on task-completion rate or direct user ratings \citep{sensitive, pred, hancock2019learning}. 
For evaluating conversational systems, metrics like task-completion rate and direct user ratings have been used \citep{sensitive, pred, hancock2019learning}. 
While these studies primarily focus on predicting offline (session-level) user satisfaction, Choi et al. in \citep{ConvSAT} proposed a neural model that predicts both online and offline satisfaction labels. 
We will use their model, ConvSAT, to define a metric to evaluate user satisfaction with different types of news recommendations.
% We will be extending the ideas proposed here to define a new set of metrics to evaluate user engagement and satisfaction with different types of news recommendations.

\section{System Description}
\label{sec: system description}
In this section, we first give an overview of the open-domain conversational agent that was used for our experiments. We then provide details of the news component of the system which is responsible for answering news-related queries and making news recommendations.
\vspace{-3mm}

\subsection{Conversational Agent Overview}
% \todo[inline]{Add high-level description of IrisBot, similar to what I added to Jason's SSML paper}
This study was performed as part of a naturalistic assessment of open-domain conversational systems, organized by the Amazon Alexa Prize Conversational AI challenge. Amazon Alexa customers were randomly assigned to each participating system, and could converse on a wide range of topics. At the end of the conversation, the customer could optionally leave a rating (1.0-5.0) and an optional comment feedback. 

Our goal was to develop a conversational agent that helps the user be informed about the world around them, while being entertained and engaged. 
Our agent incorporated real-time search, informed advice, and latest news into the conversation, by attempting to discuss and share information on relevant latest topics and opinions in the News, Sports, Entertainment, and general knowledge.
% , by accurately detecting the user's intent, both explicitly stated and implied from the conversation context. 
% The detailed description of the agent architecture, dialogue management and response ranking and generation is reported elsewhere [Citations omitted for anonymity].
The detailed description of the agent architecture, dialogue management and response ranking and generation can be found in our technical report~\citep{ahmadvand2018emory}.

% \begin{table}[htbp]
%       \centering
%           \begin{tabular}{l l l l}
%           \toprule
%           Movie & Music&News&Pets  \\ 
%           Sports&Travel&Games&Cars \\ 
%           \bottomrule
%          \end{tabular}
%       \caption{Topics handled by the system}
%       \label{tab:topiclist}
% \end{table}

The system could handle 8 major topics namely Movies, Music, News, Pets, Sports, Travel, Games and Cars.
Each of these topics was handled by a domain-specific component or "mini-skill", which would keep the user engaged in the same domain, e.g. movies, as shown in the example conversation in Table \ref{tab:example_conversation}. If the user lost interest in the topic, or the component failed to give a satisfactory response, a new topic or component would be suggested to the user. 
% The component to recommend was chosen randomly for some versions of our bot, and based on a pre-defined popularity-based sequence for others.
The same component was not recommended again until all the other components had already been recommended once to the user.
For example, if the user is first recommended a news briefing, they will not be recommended news again until they have been recommended movies, music, travel and all the other topics.
\vspace{-3mm}
\subsection{News-Oriented Conversational Agent}
\subsubsection*{Data Sources:}
The following API were used for crawling news articles and online news retrieval to provide news to the user.
\begin{itemize}
    \item {\em Microsoft Bing News search API}: \footnote{\href{https://azure.microsoft.com/en-us/services/cognitive-services/bing-news-search-api/}{https://azure.microsoft.com/en-us/services/cognitive-services/bing-news-search-api}}: Given a topic, the API returns a list of news headlines, a short snippet from the start of the article, source, and date of publication. The API also provides trending news organized by category, like sports or politics.
    \item {\em Reuters RSS Feed}: We crawl the daily news briefing from Reuters. Each story in the briefing is shortened to one or two sentences, so that the briefing can be read out to the user in about 45 seconds.
\end{itemize}

\subsubsection*{News retrieval:}
If news on a trending topic or entity is recommended and the user accepts the recommendation, we query Bing News API for news on the trending topic or entity. We then keep offering more news on the topic or entity as in the conversation in Table \ref{tab:example_conversation} till the user refuses, or changes the subject from news to something else. 
If the user accepts the News Briefing recommendation, we return the daily news briefing crawled from Reuters, and then ask the user if they want to continue talking about the news.
If the user accepts the generic news recommendation, we ask them to pick a category from sports, politics, technology and other, and then provide trending news from that category.

We describe our analysis of these methods in Section \ref{sec: results}. 
\vspace{-3mm}

\section{Experimental setup}
\label{sec:experiments}
\subsection{Dataset Overview}
Our conversational system was deployed as part of the
\href{https://developer.amazon.com/alexaprize}{Amazon Alexa Prize Challenge 2018} as mentioned in Section \ref{sec: system description}. The results are based on a total of 6994 conversations spread across June 2018 to August 2018, in which news was recommended at least once. 
\vspace{-5mm}
\subsection{Online Satisfaction Annotation}
We used the pretrained ConvSAT\citep{ConvSAT} model for predicting the probability of the user being satisfied at the end of the conversation, given the conversation upto an intermediate turn. The model proposed in the paper was trained to predict satisfaction labels for the same conversational agent which we used for conducting our experiments\citep{ahmadvand2018emory}.
As discussed in \citep{ConvSAT},
% We evaluated the model on a human-annotated dataset with 100 conversations sampled randomly from our conversation corpus. The two annotators were asked to predict satisfaction labels (binary) for the turn based on the previous turns. T
the model achieved 1.072 root mean squared error (RMSE) and 0.772 mean absolute error (MAE) on a human-annotated test dataset.

\subsection{Evaluation Metrics}
% \todo[inline]{Provide itemized list or paragraph for each metric, with names bolded, with short definition of each, equation if necessary.}

We evaluate the different formulations of news recommendations on the following four metrics.
\vspace{-2mm}

\paragraph{\textbf{Acceptance Rate}}
We define the acceptance rate of a formulation of news recommendation as the fraction of times the user explicitly accepted that type of recommendation.
\vspace{-2mm}
\paragraph{\textbf{Post-Recommendation Interruption Rate}}
This metric indicates the fraction of times the user asked to end the conversation at the turn when a recommendation of that formulation was made.
\vspace{-2mm}
\paragraph{\textbf{Post-Recommendation User Engagement}}
We define post recommendation user engagement as the average number of turns for which the user engaged with the news component after accepting that formulation of recommendation.
\vspace{-2mm}
\paragraph{\textbf{Post-Recommendation User Satisfaction}}
We used the user-satisfaction model described in \citep{ConvSAT} to measure post-recommendation user satisfaction. The model gives the probability of the user being satisfied at the end of the conversation given the conversation upto an intermediate turn. We used it to measure the probability of satisfaction at the turn when news was recommended and the turn after it. We define a formulation's post recommendation user satisfaction as the fraction of times the probability of satisfaction was higher after the recommendation than before.
\vspace{-3mm}

\section{Results and Discussion}
\label{sec: results}
We first investigate the recommendation { \em acceptance rate} (Section \ref{results:acceptance}) for different presentation conditions, which is indicative of how effective the presentation is in getting the user interested in news. Once the recommendation has been accepted, we analyze its {\em post-recommendation user engagement} (Section \ref{results:engagement}) and {\em post-recommendation user engagement satisfaction} (Section \ref{results: satisfaction}).
We then look at the {\em post-recommendation interruption rate} (Section \ref{results: interruption}) to see if the presentation can ever be detrimental to the user experience. 
% The tables show the values for all metrics for both new and returning users, and the number of samples we had for each condition.
% The improvement is reported with respect to the generic news recommendation.
% \todo[inline]{Provide short overview paragraph: We now report experimental results. First, we investigate recommendation acceptance for different presentation conditions; once a recommendation is accepted, we study the subsequent engagement (Section xx). Then, we report ... }

\subsection{Recommendation Acceptance Rate}
\label{results:acceptance}
As Table \ref{TABLE: acceptance} indicates, recommendations based on entities mentioned by the user in the current conversation have a significantly higher acceptance rate compared to other types (improvement of $97.3\%$ and $111\%$ over baseline). For returning users, the rate of acceptance of news on past entities is also significantly high. 
 Trending news recommendation, on the other hand, shows little improvement (not statistically significant) over the generic news recommendation. The reason for that is the user is unlikely to know what a trending news headline is about, and those recommendations often fail to get the user interested in news if they don't want to hear the trending news in general.
 \vspace{-2mm}

\begin{table}[H]
  \begin{tabular}{l|lr|lr}
    \toprule
    &New &$N_{new}$ & Returning & $N_{ret}$\\
    \midrule
    Entity (curr) & \textbf{0.53 (+133\%)}& 1516 & \textbf{0.49 (+97.3\%)} & 221\\
    Entity (past) & N/A& N/A & \textbf{0.53 (+111\%)} & 34\\
    News Briefing & \textbf{0.41 (+80.8\%)}& 6939 & 0.36 (+44.7\%) & 1161\\
    Trending News & 0.24 (+6.9\%)& 7202 & 0.23 (-7.2\%) & 69\\
    Generic & 0.23& 3865 & 0.25 & 52\\
  \bottomrule
\end{tabular}
\caption{Recommendation acceptance rates for all formulations for new and returning users, with relative improvements over the generic news recommendation. }
\label{TABLE: acceptance}
\end{table}
\vspace{-10mm}
\subsection{Post-Recommendation User Engagement}
\label{results:engagement}
\vspace{-3mm}
\begin{table}[H]
  \begin{tabular}{l|lr|lr}
    \toprule
    &New &$N_{new}$ & Returning & $N_{ret}$\\
    \midrule
    Entity (curr) & \textbf{1.85 (-43.5\%)}& 807 & {1.70 (-48.7\%} & 109\\
    Entity (past) & N/A& N/A & 1.55 (-53.0\%) & 18\\
    News Briefing & 1.74 (-46.7\%)& 2866 & 1.95 (-40.9\%) & 420\\
    Trending News & \textbf{1.98 (-39.37\%)}& 1759 & \textbf{1 (-69.8\%)} & 16\\
    Generic & 3.27& 883 & 3.30 & 13\\
  \bottomrule
\end{tabular}
\caption{Post-recommendation user engagement for all formulations for new and returning users, with relative improvements over the generic news recommendation.}
\label{TABLE: engagement}
\end{table}
\vspace{-5mm}
As Table \ref{TABLE: engagement} shows, generic recommendations have the highest number of follow-on turns. A possible reason for that is that users who accept the generic news recommendation are the ones who are actually interested in the news as a whole and not just specific topics. 
Also, the generic recommendation is followed by the bot asking the user to pick a category or topic of interest, which leads the user to engage for at least one more turn.

Although trending news topics have low acceptance rates, they have higher post-recommendation engagement compared to entity-based recommendations. ( p value = $1.8 * 10^{-36}$). That is because for entity-based recommendation, news component will return {\em some} recent news about the entity mentioned by the user, but the trending news is more likely to be interesting.
% This was especially the case for entities from past conversations.
% Out of the 17 users who accepted the suggestions, only 5 heard the news for more than one turn. From examining the logs for these conversations, we see that it was for entities where the first news result was recent and interesting, like Donald Trump and Casting Crowns, a band that was touring at the time. Not all recommended entities were newsworthy.
\vspace{-2mm}
\subsection{Post-Recommendation User Satisfaction}
\vspace{-2mm}
\label{results: satisfaction}
\begin{table}[H]
  \begin{tabular}{l|lr|lr}
    \toprule
    &New &$N_{new}$ & Returning & $N_{ret}$\\
    \midrule
    Entity (curr) & \textbf{0.90 (-7.4\%)}& 807 & 0.88 (-11.0\%) & 109\\
    Entity (past) & N/A& N/A & 1.0 (0.0\%) & 18\\
    News Briefing & \textbf{0.90 (-8.0\%)}& 2866 & 0.85 (-14.0\%) & 420\\
    Trending News & \textbf{0.94 (-3.54\%)}& 1759 & 0.66 (-33.3\%) & 16\\
    Generic & 0.98& 883 & 1.0 & 13\\
  \bottomrule
\end{tabular}
\caption{Post-recommendation user satisfaction for all formulations for new and returning users, with relative improvements over the generic news recommendation.}
\label{TABLE: satisfaction}
\end{table}
\vspace{-5mm}
The  generic news recommendation causes increase in immediate satisfaction the most number of times, possibly because users who accept the suggestion have prior interest in news.
% Increase in immediate satisfaction is also significantly more frequent ( p value = $1.8 * 10^{-12}$) for trending news compared to entity-based recommendation.

\subsection{Post-Recommendation Interruption Rate}
\label{results: interruption}
\begin{table}[H]
  \begin{tabular}{l|lr|lr}
    \toprule
    &New &$N_{new}$ & Returning & $N_{ret}$\\
    \midrule
    Entity (curr) & \textbf{0.03 (-51.4\%)}& 807 & 0.02 (-41.2\%) & 109\\
    Entity (past) & N/A& N/A & 0.0 (-100\%) & 18\\
    News Briefing & 0.07 (+0.1\%)& 2866 & 0.06 (+59\%) & 420\\
    Trending News & 0.07 (-5.28\%)& 1759 & 0.0 (-100\%) & 16\\
    Generic & 0.07& 883 & 0.04 & 13\\
  \bottomrule
\end{tabular}
\caption{Post-recommendation interruption rate for all formulations for new and returning users, with relative improvements over the generic news recommendation.}
\label{TABLE: interruption}
\end{table}
\vspace{-5mm}
As Table \ref{TABLE: interruption} shows, entity-based recommendation has the lowest {\em post-recommendation interruption rate}. Users very rarely end the conversation on being offered news on an entity that they expressed interest in during the current conversation.
Although News Briefing recommendation has relatively high acceptance rate, it also has the most interruption. Users generally find news briefings useful. However, because the experiments were conducted with a "socialbot" many users were not interested in a 45-second long news briefing.

\section{Conclusions and future work}
\label{sec:discussion}

In order to engage a user about news, a conversational system needs to proactively suggest news to the user first. To the best of our knowledge, this was the first study comparing the effectiveness of different options of presenting news recommendation in a conversational setting. We proposed four quantitative ways of evaluating news recommendation, specifically designed for the conversational setting, and evaluated multiple presentation methods on thousands of conversations with real users. Our results showed that there is a significant difference in how users respond to different presentation styles, with the proposed metrics emphasizing different aspects of the system performance. In particular, high acceptance rate of news recommendations on entities from a user's past conversation indicates that users retain interest in the topics they like across conversations and tend to want to hear updates on them.
Trending news is more likely to be relevant to the user compared to news on an entity. 
% (The entity itself might be more relevant to the user because they talked about it, but there might not be any recent news about it.) 
But acceptance is higher for entity-based recommendations. That suggests that the representation affects acceptance in addition to the quality of recommendations.
In future work, we plan to investigate how to better identify the news items to recommend based on the user's conversational history and long term profile, and how to engage the user in a conversation about a news topic after the initial news item was accepted. Together, our results, evaluation metrics, and analysis provide valuable insights into improving conversational news recommendation, and more generally information-oriented conversational agents.

\bibliographystyle{ACM-Reference-Format}
\bibliography{bibliography}

\balance
\end{document}